\documentclass[prl,twocolumn,showpacs,superscriptaddress,aps]{revtex4-1}
\usepackage{graphicx,amssymb,amsmath,psfrag,color}
\usepackage[colorlinks=true,citecolor=blue,linkcolor=blue,urlcolor=blue]{hyperref}
\begin{document}
\definecolor{darkgreen}{rgb}{0,0.5,0}
\newcommand{\be}{\begin{equation}}
\newcommand{\ee}{\end{equation}}
\newcommand{\jav}[1]{#1}

\title{Out-of-time-ordered density correlators in Luttinger liquids}

\author{Bal\'azs D\'ora}
\email{dora@eik.bme.hu}
\affiliation{Department of Theoretical Physics and MTA-BME Lend\"{u}let Spintronics 
Research Group (PROSPIN), Budapest University of Technology and Economics, 1521 Budapest, Hungary}
\author{Roderich Moessner}
\affiliation{Max-Planck-Institut f\"ur Physik komplexer Systeme, 01187 Dresden, Germany}

\date{\today}

\begin{abstract}
Information scrambling  and the butterfly effect in chaotic quantum systems can be diagnosed by out-of-time-ordered  (OTO) commutators through
an exponential growth and large late time value. We show that the latter feature shows up in a strongly correlated many-body system,  a Luttinger liquid, 
whose density fluctuations we study at long and short wavelengths,  
both  in equilibrium and after a quantum quench. We find rich behaviour combining robustly  universal and non-universal features.
The OTO commutators display temperature and initial state independent behaviour, and grow as $t^2$ for short times.
For the short wavelength density operator, they reach a sizeable value after the light cone \jav{only} in an interacting Luttinger liquid, \jav{where the bare excitations break up 
into collective modes.}
We benchmark our findings numerically on an interacting spinless fermion model in 1D, and find persistence of central features
even in the non-integrable case. As a non-universal feature, the short time growth exhibits a distance dependent power.
\end{abstract}


\maketitle

\paragraph{Introduction.}
Spectacular experimental progress in the study of coherent quantum
dynamics has focused much attention on the question how many-body
 systems evolve in real time\cite{polkovnikovrmp,dziarmagareview}. 
Particularly interesting is the study of ergodic versus non-ergodic behaviour. This builds on previous
work under the heading of classical and quantum chaos, where ideas
such as the Lyapunov exponents (and the concomitant 'butterfly
effect') have turned out to be useful, \jav{together with level statistics} of the Hamiltonian. 
This has been supplemented by questions about how information -- in the
form of correlations or entanglement -- spreads\cite{liebrobinson,cheneau,wenho,nahum}. Here, the concept of scrambling
encodes the loss of information under time evolution, in
particular asking the question to what extent different initial
states can be distinguished at later times\cite{roberts,roberts2016}.

Recently, out-of-time-ordered (OTO) correlation functions\cite{larkin} have been identified as quantities \jav{providing
 insights into quantum chaos and information scrambling}.
The OTO commutator is defined as
\begin{gather}
C(t)=\left\langle \left[V^{},W(t)\right] \left[W^+(t),V^+\right]\right\rangle\geq 0,
\label{ct}
\end{gather}
where $V$ and $W$ are usually local operators, possibly separated by some spatial distance
and $W(t)=\exp(iHt)W\exp(-iHt)$.
It contains terms of the form $-\langle VW(t)V^+W^+(t)\rangle$, coined  OTO correlator due to its unusual  temporal structure.
It probes the spread of information, in particular signaling the presence
of quantum chaos, with a growth bounded by a thermal Lyapunov
exponent\cite{maldacena2016}. 
Much effort, including experiments\cite{junli,garttner,nyao,swingle,zhugrover,campisi,aleiner}, 
has been devoted to its study, with intriguing connections to black hole physics and
random matrix theory\cite{maldacena2016,cotler} appearing, while it has turned
out that a simple 'mesoscopic', Sachdev-Ye-Kitaev model\cite{sachdevye,kitaev,maldacenaprd,larkin,fritzpatrick} 
captures many interesting phenomena.

\jav{The quantum butterfly effect (sensitivity to small perturbations), occurs in chaotic systems. 
It is diagnosed by the OTO commutator
via an exponential growth before becoming of order $2\langle VV^+\rangle\langle WW^+\rangle$ at late times\cite{roberts,roberts2016,maldacena2016}.
While this in itself is an interesting phenomenon, it is equally important to determine what universal features
characterize OTO commutators in well-studied yet non-trivial models of
condensed matter, even though these are expected not to saturate any 
 scrambling bounds as they reside on finite-dimensional lattices.}

In order to advance this program, we consider Luttinger liquids (LLs), \jav{realized in a variety of settings\cite{giamarchi,cazalillarmp}.
LLs describe the low energy physics of both integrable (i.e.\ non-ergodic) and non-integrable (ergodic) 
critical\cite{sachdev} 1D systems, predicting universal behaviour  for the long time dynamics,  
irrespective of the (non-)integrability of the system. 
LLs thus represent an ideal setting to gain universal information about the quantum butterfly effect
and to disentangle chaotic from regular behaviour.
Moreover, in the presence of interactions, the original non-interacting quasiparticle description breaks down 
as excitations fractionalize into collective 
bosonic modes. This in particular turns a Fermi gas into  a  non-Fermi liquid. 
How these effects combine in the OTO commutator in LLs is the main subject of our work.}

Here, we focus on Eq.~\eqref{ct} in LLs~\footnote{We also consider a generalized OTO commutator  without hermitian conjugation, which also reflects the phase fluctuations of the commutator.}
and find that in the  OTO density commutator,
an initial rise $\propto t^2$  
 builds up to a strong signal
upon the arrival of the light cone, beyond which it saturates. 
\jav{This saturation is reminiscent to the quantum butterfly 
effect in chaotic models. However,  we suspect that it occurs here 
due to the replacement of the original excitations by new bosonic collective excitations.
Indeed, in the non-interacting case, both  features are absent together.}
The saturation is due to short-wavelength
degrees of freedom, while the long-wavelength contribution -- which exhibits a large degree of universality -- 
vanishes at large times.
Our findings are insensitive to the choice of  initial thermal or ground states, 
as well as whether we time-evolve an eigenstate or follow a quantum quench protocols\cite{polkovnikovrmp,dziarmagareview}.
These central features are reproduced by exact diagonalization studies of
the XXZ model.
Remarkably, the short-time $t^2$ behaviour gives way to a $t^{2x}$ rise for spatially separated local densities (with separation $x$), 
reflecting the prominent role of  microscopic details in the model under study.

\paragraph{Luttinger model.} 
The low energy description of LLs is   in terms of bosonic sound-like collective excitations
with  Hamiltonian\cite{giamarchi,delft}
\begin{equation}
H=\sum_{q\neq 0}  v|q| b_q^{+} b^{\phantom{+}}_q
+\frac{g(q)}{2}[b^{\phantom{+}}_qb^{\phantom{+}}_{-q}+b_q^{+} b_{-q}^{+} ],
\label{ham0}
\end{equation}
where $b_q$ is the annihilation operator of a bosonic density wave, $g(q)=g_2|q|$, with $g_2$  the interaction strength, and $v$ the sound velocity of the non-interacting system.
The interaction is also characterized by the dimensionless Luttinger parameter, 
 $K=\sqrt{(v-g_2)/(v+g_2)}$.
Eq. \eqref{ham0} is diagonalized by a Bogoliubov rotation, and the dispersion relation is $\omega_q=v_f|q|$
with the renormalized velocity $v_f=\sqrt{v^2-g_2^2}$.
The transformation gives $b_q=\cosh(\theta)B_q+\sinh(\theta)B^+_{-q}$ 
with $B_q$ the new boson operators, $\cosh(\theta)=\frac{K+1}{2\sqrt K}$ and $\sinh(\theta)=\frac{K-1}{2\sqrt K}$, which are $q$ independent for the present case.

The time dependence of the original boson field is
\begin{gather}
b_q(t)=u_q(t)b_q+v_q(t)^*b^+_{-q}
\label{bt}
\end{gather}
with 
$u_q(t)=\cos(\omega_q t)-i\sin(\omega_qt)\cosh(2\theta)$ and $v_q(t)=-\sin(\omega_qt)\sinh(2\theta)$.
Any expectation value is taken using a Bogoliubov rotation into the $B_q$ basis. One can consider both a quantum quench and equilibrium time evolution.
In the former case, the LL parameter $K$ in the time evolution, Eq. \eqref{bt} differs  from those in the $B_q$ operators. For the sake of simplicity, we assume quenching from a non-interacting, $K=1$ state, therefore
the $B_q$ operators are identical to the $b_q$'s. In the equilibrium case, the same LL parameter is used  for both  time evolution and Bogoliubov rotation.

We focus on the long wavelength ($q\sim0$) density fluctuations and a vertex operator\cite{delft}. Their bosonized versions are
\begin{gather}
n_0(x)=-\frac 1\pi \frac{\partial\phi(x)}{\partial x}, \hspace*{5mm}
V_n(x)=\exp(in\phi(x)),\label{ndw}
\end{gather}
respectively, where $n$ is integer and $V_{\pm 2}$ corresponds to the $\pm 2k_F$ Fourier component of the 
short wavelength density operator with $k_F$ the
Fermi wavenumber. This operator is responsible for e.g. the density wave phase transitions\cite{gruner,giamarchi,nersesyan}
and Friedel oscillations.
A similar expression describes the phase fluctuations in a 1D quasi condensate\cite{cazalillarmp,giamarchi}.
The bosonic field in Eqs. \eqref{ndw} is expressed in terms of the canonical bose operators as
\begin{gather}
\phi(x)=\sum_{q\neq 0}\sqrt\frac{\pi}{2|q|L}\left(\exp(iqx)b_q+\textmd{h.c.}\right).
\label{phi}
\end{gather}

\paragraph{Long wavelength density.}
The OTO commutator  of the long wavelength field builds on the density response function 
$\chi(t)=\langle\left[n_0(x,t),n_0(0,0)\right]\rangle$,
where the bare commutator, $\left[n_0(x,t),n_0(0,0)\right]$ is already a c-number\cite{giamarchi,imambekov}
\jav{due to the linear dispersion in Eq. \eqref{ham0}}, therefore 
$\chi(t)=\frac{iK}{\pi^2}\left(\frac{\alpha(v_ft+x)}{(\alpha^2+(v_ft+x)^2)^2}
+(x\rightarrow -x)\right)$  with $\alpha$ the short distance cutoff\cite{giamarchi}.
The OTO commutator then yields
\begin{gather}
C_0(t)=\left\langle\left|\left[n_0(x,t),n_0(0,0)\right]\right|^2\right\rangle=|\chi(t)|^2.
\label{c0}
\end{gather}
It has several interesting consequences: a.) from Eq. \eqref{c0} and the fact that $\chi(t)$ does not contain the OTO correlator,
this OTO commutator is not influenced by the OTO correlator either.
b.) since the bare commutator is already a temperature independent c-number, the expectation value in
$C_0(t)$
becomes independent of both temperature\cite{maldacena2016} and the wavefunctions--it only depends on the time evolution operator \jav{ 
and is completely independent of the
initial state.}
Therefore, there is no distinction between a quantum quench and equilibrium evolution.
c.) Putting all this together, $C_0(t)$
 grows in a $t^2$ manner initially, exhibits double peaks at around the light cone and decays as $1/t^6$ for long times.
The $t^2$  is the lowest possible power at short times  of the OTO commutator,
 with prefactor  $\langle| \left[\left[H,W\right],V\right]|^2\rangle$,
unless this expectation value vanishes.
As we show later for  the \emph{short} wavelength density fluctuations, this can also occur.

Note that the Fourier transform of $\chi(t)$ gives 
the  dynamical structure factor, $S(\omega,q)\sim K|q|\delta(\omega-v_f|q|)$ in a LL, indicating that bosonic excitations  have an infinite lifetime
due to the linearized dispersion\cite{imambekov}.
In a nonlinear LL picture with finite curvature from the non-interacting band structure\cite{samokhin}, 
the correlation function changes and develops additional tails. How
curvature and other higher energy features are manifested in the OTO commutator is an intriguing open question, but is beyond the scope of the present work.
~\footnote{We have also investigated the case of time evolution with a gapped Hamiltonian\cite{iuccisinegordon}. In this case, 
the OTO commutator decays as
$\Delta\cos^2(\Delta t)/t^{3}$. 
Similar temperature independent and power law decaying commutators have also been reported for the  Falicov-Kimball  model\cite{tsuji}.}

\paragraph{Short wavelength density.}
The OTO commutator of the general vertex operator is more involved. The simple commutator is rewritten\cite{giamarchi,delft}, as
\begin{gather}
\left[V_n(x_1),V_{-m}(x_2)\right]=2e^{i(n\phi_1-m\phi_2)}\sinh\left(\frac{nm}{2}\left[\phi_1,\phi_2\right]\right)
\label{commshort}
\end{gather}
with $\phi_{1,2}=\phi(x_{1,2})$ \jav{and the commutator in Eq. \eqref{commshort} yields a vertex-like operator, in contrast to the long wavelength case}. The commutator of the $\phi$ fields is a temperature independent c-number and is evaluated as
\begin{gather}
\left[\phi(x,t),\phi(0,0)\right]=-i\frac{K}{2}\arctan\left(\frac{v_ft+x}{\alpha}\right)+(x\rightarrow -x).
\label{commdx}
\end{gather}

Now, we calculate  the OTO commutator from Eq.~\eqref{commshort}.
The two exponential fields are hermitian conjugates of each other, $\exp(i(n\phi_1-m\phi_2))\exp(i(m\phi_2-n\phi_1))$, and need to be contracted. 
In the properly regularized theory, the exponential fields are point split\cite{delft} and the exponents then merged. After taking the expectation value, we get 1, 
independent of temperature. Putting all this together, only the commutator in Eq.~\eqref{commdx} remains and
\begin{gather}
C_{dw}(t)=\left\langle \left|\left[V_2(x,t),V_{-2}(0,0)\right]\right|^2\right\rangle=\nonumber\\
=4\sin^2\left(K\arctan\left(\frac{v_ft+x}{\alpha}\right)+\left( x\rightarrow -x\right) \right)\nonumber \\
\xrightarrow{t\rightarrow\infty}
4\sin^2(K\pi).
\label{abs2}
\end{gather}
This expression grows with $t^2$ for $v_ft\ll |x|$, and rises sharply to $4\sin^2(K\pi)$ on hitting the light cone. Its long time value vanishes in 
the non-interacting limit~\footnote{For $K=1$, it simplifies to $16(\alpha v_ft)^2/((v_ft+x)^2+\alpha^2)/((v_ft-x)^2+\alpha^2)$, rising sharply with time and decaying as $1/t^2$.}
but can give a sizable contribution to the commutator for a range of $K$ values. 
\jav{Surprisingly, $C_{dw}(t)\leq 4$ follows from the Cauchy-Schwarz inequality ($\| ab\|\leq \|a\|\|b\|$) and $\| V_n(x)\|=1$ and  its maximal value from Eq. \eqref{abs2} 
is reached at late times for e.g.  $K=1/2$. Note that even in a suitably chaotic system, the late time value is expected to be 2\cite{maldacena2016}. }
This is in sharp contrast to the expectation value of the simple commutator 
in Eq. \eqref{commshort}, which, for $n=m=2$, gives the retarded short wavelength 
charge susceptibility in a LL and vanishes in a power law fashion in the long time limit\cite{giamarchi}.
\jav{Eq. \eqref{abs2}  is  thus strongly influenced by the OTO propagator as it differs from the square of the short wavelength charge susceptibility.
In getting the $C_{dw}(t)$ OTO commutator, Eq. \eqref{commshort} itself is a vertex operator like quantity, and becomes
 independent of both temperature and initial state after taking its expectation value.}
Even though the Luttinger model is not chaotic, this commutator still exhibits
some characteristics of the butterfly effect in the sense that the late time limit of the OTO commutator reaches a sizeable value, indicating the effect of the OTO propagator.

Finally, the simple square of Eq. \eqref{commshort} includes  phase information into the
OTO commutator, which 'degrades' the signal at long times, in particular replacing
the above saturation  with a decaying form sensitive to
temperature (power-law versus exponential) as well as evolution
protocol\cite{EPAPS}.


\paragraph{OTO density commutator for interacting fermions.}
In order to test our results on the OTO commutator, we have studied one dimensional spinless fermions in a tight-binding chain with nearest neighbour repulsion at half filling and periodic boundary 
condition using exact diagonalization (ED). This problem is
equivalent to the 1D Heisenberg XXZ chain after a Jordan-Wigner transformation\cite{giamarchi,nersesyan}.
The Hamiltonian is
\begin{gather}
H=\sum_{m=1}^N \frac{J}{2} \left(c^+_{m+1}c_m +\textmd{h.c.}\right)+J_z n_{m+1}n_m,
\label{xxz}
\end{gather}
where $c$'s are fermionic operators and the $c_{N+1}=c_1$, and $J_z$ denotes the nearest neighbour repulsion. This model realizes a LL for $J_z<J$ with
 LL parameter $K=\pi/2[\pi-\arccos(J_z/J)]$.
We have evaluated the OTO commutator of the local charge density, $n_1=c_1^+c_1$ and its time evolved counterpart to be able to access directly the 
late time behaviour after the light cone.
System sizes up to $N=22$ are considered, the number of electrons being $N/2$.

\begin{figure}[h!]
\includegraphics[width=7cm]{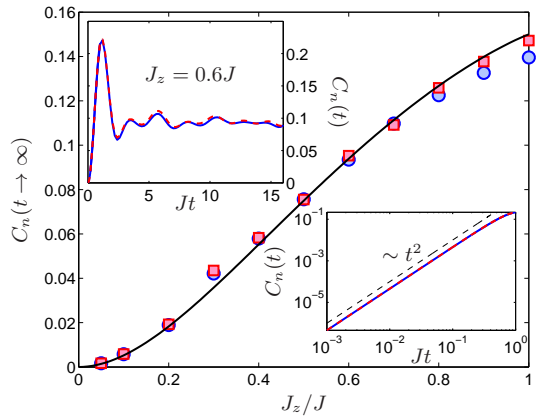}
\caption{ED result for $N=22$ for the OTO commutator for the local charge density in the interacting spinless fermion model. Red squares and blue circles denote
the late time limit in equilibrium and after a sudden quench from the non-interacting limit, respectively.
The solid line is $ f \sin(K\pi)^2$ from Eq. \eqref{abs2} with  only the  parameter $f=0.15$ adjustable.
Top inset:  numerical time evolution of the OTO commutator  with the representative value $J_z/J=0.6$ in equilibrium (red dashed line) and after sudden quench (blue solid line) from the non-interacting case, $J_z=0$.
For short time ($t<1/J$), it follows the predicted $t^2$ (bottom inset).
\label{nnll}}
\end{figure}

According to Ref. \onlinecite{maldacena2016}, this should approach $2\langle n_1\rangle^2=2\left(\frac 12\right)^2=\frac 12$ at late times when the butterfly effect occurs.
In suitably chosen chaotic systems, this occurs through  the exponential growth of the OTO correlator of local operators 
under time evolution as is the case in the Sachdev-Ye-Kitaev model\cite{sachdevye,kitaev,maldacenaprd,larkin,fritzpatrick}.
In bosonized form, the local charge density\cite{giamarchi} is  $n(0)=-\partial_x \phi(0)/\pi+n_{2k_F}\cos(2\phi(0))$, where $n_{2k_F}$ depends on the short range properties
of Eq.~\eqref{xxz} available only from its exact solution\cite{lukyanov}. The OTO commutator contains both  short and long wavelength operators, but its long time limit
will be dominated by Eq.~\eqref{abs2}, i.e.\ it should approach a non-zero, constant value. 

The numerical evaluation of its OTO commutator is shown in Fig.~\ref{nnll}, 
together with the predicted behaviour. We investigate both equilibrium time evolution from an interacting ground state as well as quantum quenches from a non-interacting, $J_z=0$ ground state to
an interacting system.
We find very satisfactory agreement between bosonization and ED. In particular, a.) both  equilibrium and sudden quench OTO commutators stay very close to each other b.) 
the commutator reaches a time independent value after a transient time, c.)
the short time behaviour is $t^2$. 
The agreement on the steady state for the OTO commutator with Eq. \eqref{abs2} is surprisingly good, given it  contains the unknown prefactor $n_{2k_F}$, which can in principle also depend\cite{lukyanov} on $J_z$.
The OTO commutator is indeed of the same order as expected for the quantum butterfly effect \jav{and reaches its maximal value at $J_z=J$ ($\Leftrightarrow K=1/2$) from Eq. \eqref{abs2}}, even though the model in Eq. \eqref{xxz} is not chaotic but integrable. 
\jav{Any finite $J_z$ in Eq. \eqref{xxz} destroys the Fermi gas and induces  non-Fermi liquid (LL) behaviour, and the bare fermionic excitations do not persist but give way to collective bosonic modes.
The density operator in OTO commutator 
naturally decomposes into collective  modes during the time evolution, which could explain the large late time value.
As evident from Fig.~\ref{nnll}, absent fractionalization ($J_z=0$) implies $C_n(t\rightarrow \infty)\rightarrow 0$.}

\begin{figure}[h!]
\includegraphics[width=6cm]{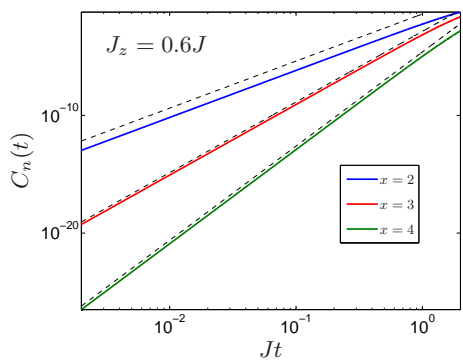}
\caption{ED result for $N=10$ for the OTO
commutator between the local charge densities $n_1$ and $n_{1+x}$ in the interacting spinless fermion model with $x=2$ (blue), 3 (red) and 4 (green) 
in equilibrium for $J_z/J=0.6$, the quench data is indistinguishable in this
time window. The thin black dashed lines denote $t^{2x}/(2x)!$.
\label{nnllx}}
\end{figure}

We have also investigated the OTO commutator between $n_1$ and $n_{1+x}$ with $x$ positive integer. The late time behaviour after hitting 
the light cone agrees with our previous findings and takes an order one value as in Eq. \eqref{abs2}.
For shorter times, on the other hand, the OTO commutator grows as $t^{2x}/(2x)!$, shown in Fig. \ref{nnllx}.
This follows from a Baker-Campbell-Hausdorff expansion of the $W(t)$ in Eq. \eqref{ct}, with the nested commutators\cite{roberts2016} 
\begin{gather}
W(t)=W+it\left[H,W\right]+\frac{(it)^2}{2!}\left[H,\left[H,W\right]\right]+\dots.
\label{wt}
\end{gather}
For $W=n_{x+1}$,  the coefficient of the $t^2$ term in the OTO commutator arises 
from the second term in Eq. \eqref{wt} as $\langle \left[\left[H,n_{x+1}\right],n_{1}\right]^2\rangle$. However, this 
vanishes for $x>1$ since the $\left[H,n_{x+1}\right]$ commutator contains
fermionic operators with indices from $x$ to $x+2$, which commute with $n_1$.  At distance $x$, the 
 commutators start contributing at $x$th order nesting,
yielding a leading power $t^{2x}$.

This illustrates that the short time growth of the OTO commutator is dictated by the short range properties of the model -- not accounted
for by the low-energy theory captured by bosonization -- such as the range of the hopping processes or interactions.
The short time growth, before reaching the light cone, is thus seen to 
depend on the high energy (ultraviolet) part of the spectrum as well\cite{roberts2016}.
We emphasize that for $x=0$, the $t^2$ prediction is also confirmed numerically, shown in Fig. \ref{nnll}.

The XXZ Heisenberg chain in Eq. \eqref{xxz} is integrable but contains high energy features not    
accounted for by the LL model. These dominate only the transient response around $tJ\sim 1$ in Fig. \ref{nnll}.
Integrability is destroyed by adding a second nearest-neighbour  density-density (i.e. $J_z^\prime \sum_{m} n_{m+2}n_m$),
or Ising interaction\cite{hallberg} in the spin language, which  we have also studied numerically. 
It reproduces the central features found for the integrable case, in particular the $t^2$ initial growth of
the OTO commutator of the $n_1$ local density as well as the saturation of the OTO commutator after the light cone
~\footnote{The LL picture does not predict any temperature dependence in the OTO density commutators. 
In non-integrable lattice models, we cannot exclude the possibility of (temperature dependent) Lyapunov exponent, which, then,
does not originate from the universal low energy degrees of freedom in the Luttinger model 
but from the miscoscopic details of the lattice model, analogously to the short time
growth in Fig. \ref{nnllx}.}.

\paragraph{Concluding remarks.}
We have investigated the OTO correlator  
\jav{in equilibrium and after a quantum quench in one of the canonical low-dimensional model systems in the thermodynamic limit,  i.e. in a Luttinger liquid.}
The OTO commutators display robust behaviour, independent of temperature, initial state and protocol (equilibrium time evolution vs.\ quantum quenches).

In general, the quantum butterfly effect is manifest in an exponential growth and a large late time value of the OTO commutator\cite{roberts,roberts2016}. 
Based on our results, the first feature is not realized\cite{fine} during the time evolution in LL.
The short time OTO commutator grows with $t^2$, i.e. the lowest possible power, 
before reaching the light cone as opposed to the exponential growth in chaotic systems.
LLs are thus slow information scramblers, with information encoded in local operators lost slowly.

Surprisingly, the second feature of the butterfly effect appears also in LLs:
the OTO commutator of vertex operators, which incorporate the phase fluctuation in a quasi-condensate and the $2k_F$ density fluctuations,
are often enhanced significantly in a strongly interacting theory after hitting the light cone.
The time at which this enhancement occurs defines the scrambling time, which, in our case, is simply the position of the light cone, i.e. $t=x/v_f$.
\jav{This large late time value, growing with interaction, occurs probably due to the replacement  of the original quasiparticles 
 by collective bosonic modes, being absent in the non-interacting case. It will be interesting to investigate 
 what the minimal
 ingredients, and alternative settings, are for such transmutation of the underlying degrees of freedom to yield a characteristic
 OTO commutator signal.}

Many of these features are benchmarked by exact diagonalization calculations on an interacting spinless fermion model, and they are remarkably robust.
We also find that the short time dynamics is extremely sensitive to the  microscopic details of the model under study,
yielding a much steeper initial growth with a power,  $t^{2x}$, depending linearly on the distance, $x$, of the densities in the OTO commutator.

\begin{acknowledgments}

We thank M. Kormos, I. Lovas and G. Tak\'acs for illuminating discussions.
This research is supported by the National Research, Development and Innovation Office - NKFIH  K105149, K108676, SNN118028 and K119442 and by the  DFG via SFB 1143.
\end{acknowledgments}

\bibliographystyle{apsrev}
\bibliography{wboson1}

\end{document}